\documentclass[12pt, epsfig]{article}                            
\usepackage{epsfig}          
\usepackage{amssymb}     
\usepackage{amsfonts}    
\newskip\humongous \humongous=0pt plus 1000pt minus 1000pt

\newif\ifdtup

\catcode`\@=11
 
\@addtoreset{equation}{section}
\def\theequation{\arabic{equation}}
 
\def\@normalsize{\@setsize\normalsize{15pt}\xiipt\@xiipt
\abovedisplayskip 14pt plus3pt minus3pt%
\belowdisplayskip \abovedisplayskip
\abovedisplayshortskip \z@ plus3pt%
\belowdisplayshortskip 7pt plus3.5pt minus0pt}
 
\def\small{\@setsize\small{13.6pt}\xipt\@xipt
\abovedisplayskip 13pt plus3pt minus3pt%
\belowdisplayskip \abovedisplayskip
\abovedisplayshortskip \z@ plus3pt%
\belowdisplayshortskip 7pt plus3.5pt minus0pt
\def\@listi{\parsep 4.5pt plus 2pt minus 1pt
     \itemsep \parsep
     \topsep 9pt plus 3pt minus 3pt}}
 
\relax

\catcode`@=12
 
\catcode`\@=11
 
\def\section{\@startsection{section}{1}{\z@}{3.5ex plus 1ex minus
   .2ex}{2.3ex plus .2ex}{\large\bf}}
 
\def\thesection{\arabic{section}.}

\def\appendix{\setcounter{section}{0}
 \def\thesection{Appendix \Alph{section}:}
 \def\theequation{\Alph{section}.\arabic{equation}}}
 
\def\YGrule{0.4}   
\def\YGbox{6.5}    
\def\SymBoxes#1#2#3#4{\newdimen\un@t \un@t#3%
\raisebox{#1}{\rule{#2\un@t}{#4}\hskip-#2\un@t
\@tempdimb\un@t \advance\@tempdimb by-#4\@tempcntb#2\relax%
\@whilenum{\@tempcntb>0}\do{
\rule{#4}{\un@t}\hskip\@tempdimb \advance\@tempcntb by\m@ne}%
\hskip-#2\un@t \rule[\un@t]{#2\un@t}{#4}%
\rule[\un@t]{#4}{#4}\hskip-#4
\rule{#4}{\un@t}}\hskip-#4}                
\def\Young{\@ifnextchar[{\@Young}{\@Young[0]}}
\def\@Young[#1]#2{\newdimen\YG@unit \YG@unit\YGbox pt%
\newdimen\h@ight \h@ight#1\YG@unit \@tempcnta-1\relax
\@tfor\c@ount:=#2\do{\advance\@tempcnta by\@ne}
\@tempdima\@tempcnta\YG@unit%
\advance\h@ight by\@tempdima\relax     
\@tfor\c@ount:=#2\do{\SymBoxes{\h@ight}{\c@ount}{\YG@unit}{\YGrule pt}%
\@tempdima-\c@ount\YG@unit \hskip\@tempdima%
\advance \h@ight by -\YG@unit}         
\@tempdima\YG@unit \multiply\@tempdima by\@car#2\@nil %
\hskip\@tempdima}                      
\def\YoungTab{\@ifnextchar[{\@YoungIdx}{\@YoungIdx[0]}}
\def\@YoungIdx[#1]{\@ifnextchar[{\@iYoungIdx[#1]}{\@iYoungIdx[#1][\@empty]}}
\def\@iYoungIdx[#1][#2]#3{%
\newdimen\YG@unit \YG@unit\YGbox pt\newdimen\YG@rule \YG@rule \YGrule pt
\newcount\c@ount \c@ount\z@ \newdimen\skip@wd \unitlength\@ne pt
\newdimen\h@ight \h@ight#1\YG@unit \@tempcnta\m@ne\relax
\@tfor\d@um:=#3\do{\advance\@tempcnta by\@ne}
\@tempdima\@tempcnta\YG@unit%
\advance\h@ight by\@tempdima\relax
\@tfor\@idxlist:=#3\do{
\@tempcnta\z@\hskip.5\YG@rule\relax 
\@for\@idx:=\@idxlist\do{
\raisebox{\h@ight}{\makebox(\YGbox,\YGbox){#2$\@idx$}}
\advance\@tempcnta by\@ne}\hskip-.5\YG@rule%
\@tempdima-\@tempcnta\YG@unit \hskip\@tempdima%
\ifnum\c@ount=\z@ \skip@wd-\@tempdima\fi \relax
\SymBoxes{\h@ight}{\@tempcnta}{\YG@unit}{\YG@rule}%
\hskip\@tempdima \advance\h@ight by -\YG@unit
\advance\c@ount by\@ne}
\hskip\skip@wd}                      

\begin{document}

\newcommand{\beq}{\begin{equation}}
\newcommand{\eeq}{\end{equation}}
\newcommand{\bea}{\begin{eqnarray}}
\newcommand{\eea}{\end{eqnarray}}
\newcommand{\beas}{\begin{eqnarray*}}
\newcommand{\eeas}{\end{eqnarray*}}
\newcommand{\defi}{\stackrel{\rm def}{=}}
\newcommand{\non}{\nonumber}
\newcommand{\bquo}{\begin{quote}}
\newcommand{\enqu}{\end{quote}}
\def\de{\partial}
\def\Tr{ \hbox{\rm Tr}}
\def\const{\hbox {\rm const.}}
\def\o{\over}
\def\im{\hbox{\rm Im}}
\def\re{\hbox{\rm Re}}
\def\bra{\langle}\def\ket{\rangle}
\def\Arg{\hbox {\rm Arg}}
\def\Re{\hbox {\rm Re}}
\def\Im{\hbox {\rm Im}}
\def\diag{\hbox{\rm diag}}
\def\longvert{{\rule[-2mm]{0.1mm}{7mm}}\,}

\bigskip

\begin{titlepage}
{\hfill     IFUP-TH 16/2002} 
\bigskip
\bigskip

\bigskip
\bigskip

\begin{center}
{\Large  {\bf 
  Non-universal corrections to the tension ratios in  softly  broken    ${\cal N}=2$   $SU(N)$    gauge theory
} }
\end{center}
\vspace{1em}
\begin{center}
{\large Roberto AUZZI $^{(1,3)}$  and  Kenichi  KONISHI $^{(2,3)}$ }
\end{center}     
\vspace{1em}
\begin{center}
{\it
 Scuola Normale Superiore - Pisa $^{(1)}$   \\
Dipartimento di Fisica   ``E. Fermi"  -- Universit\`a di Pisa $^{(2)}$   \\
Istituto Nazionale di Fisica Nucleare -- Sezione di Pisa $^{(3)}$      }
\\
{\it Via Buonarroti, 2,   Ed. C, 56127  Pisa, Italy} \\
{\it E-mail: konishi@df.unipi.it,   r.auzzi@sns.it}
\end {center}

\vspace{3em}
\noindent  
{\bf Abstract:}

 {  Calculation by Douglas and Shenker of  the tension ratios  for vortices of different 
$N$-alities  in the softly broken ${\cal N}=2$  supersymmetric $SU(N)$  Yang-Mills theory,   is carried to the second order in 
 the   adjoint multiplet  mass    $m.$   
Corrections  to the ratios violating the well-known  sine  formula  are  found,  showing that it  is not a universal 
quantity.      
 }

\vspace{1.5em}

\end{titlepage}

\bigskip

\noindent {\bf 1.}      Recently  the tension ratios among the confining  vortices corresponding to sources of  different
$N$-alities  in $SU(N)$ gauge theories  have   been the subject of  some  attention,  as a quantity characterizing quantitatively 
the confining phase of these systems.     After an
interesting  suggestion from MQCD   that such ratios might  have universal  values \cite{STR},
\beq  { T_k  \o T_1}  ={ \sin {\pi k \o N}  \o  \sin {\pi  \o N}},    \label{sineform}  \eeq
a more recent study  in string theory   based on supergravity duals \cite{Kleb},  gave  model dependent 
results for  two  
${\cal N}=1$  SQCD-like theories.    
The result of direct measurement  in the  lattice (non-supersymmetric)  $SU(N)$   gauge theories
is consistent with  Eq.(\ref{sineform}) \cite{LT,Pisa}. 

Derivation  of formula such as Eq.(\ref{sineform})  in the standard, continuous $SU(N)$   gauge theories   still defies us.    The first field-theoretic 
result  on this issue was obtained by   Douglas and Shenker \cite{DS},    
  in the  ${\cal N}=2$ supersymmetric  $SU(N)$  pure Yang Mills theory, with supersymmetry   softly broken to  ${\cal N}=1$  by a small  
adjoint scalar multiplet mass  $m$. 
They  found  Eq.(\ref{sineform})  for the  ratios
of   the   tensions   of    abelian    (Abrikosov-Nielsen-Olesen)  vortices corresponding to   different  $U(1)$ factors of
the low-energy  effective (magnetic)  
$U(1)^{N-1}$   theory.    

 The  $n$-th  color component of the quark has  charges 
\beq       \delta_{n, k} - \delta_{n, k +1}, \qquad   (k=1,2,\ldots, N-1; \, n=1,2,\ldots, N) 
\eeq
 with   respect to the various  electric $U_k(1)$  gauge groups.       The source   of the 
$k$-th   ANO string thus  corresponds to the $N$-ality $k$   multiquark state,
$|k \ket  = |q_1 q_2,\ldots q_k \ket$, allowing a re-interpretation of Eq.(\ref{sineform})  as referring to the
ratio of the tension for different $N$-ality  confining strings \cite{STRASS}.   
  
   However,    physics of  the softly broken  ${\cal N}=2$  $SU(N)$  pure Yang-Mills theory is   quite different from
what  is expected in QCD.     Dynamical   $SU(N) \to U(1)^{N-1}$ breaking  introduces multiple
of   meson Regge trajectories with different slopes at low masses \cite{STRASS,YUNG}, a feature which is neither  seen in Nature nor expected in
QCD.\footnote{In fact, the same  problem is  expected  in any confining vacuum in which  such a dynamical breaking takes
place.   't Hooft's original suggestion for QCD ground state \cite{TM} is of this type.   }    For instance, another
$N$-ality
$k$ state
$|k\ket ^{\prime}= |q_2 q_3,\ldots q_{k+1} \ket$ acts as  source of the  $U_{k+1}(1)$ vortex and as the  sink of the
$U_2(1)$ vortex, which together bind $ |k \ket^{\prime}$-  anti $  |{k} \ket^{\prime}$ states with   a  tension different from $T_k$. 
The  Douglas-Shenker prediction is, so to speak, a good  prediction for a wrong theory! 
Only in the limit of ${\cal   
N}=1$ does one expect  to find one  stable vortex for each $N$-ality,   corresponding to the conserved $Z_N$ charges \cite{STRASS}. 

Within the softly broken ${\cal N}=2$  $SU(N)$ theory,   the two regimes can  be in principle smoothly interpolated
by varying the adjoint mass $m $ from zero to  infinity, adjusting appropriately $\Lambda$.  At small $m $  one has a good
local description of the low-energy effective  dual, magnetic  $U(1)^{N-1}$   theory.      The transition towards large 
$m$   regime involves both perturbative and  nonperturbative effects.    Perturbatively, there are higher  corrections
due to the ${\cal N}=1$ perturbation, $m   \, \Tr \Phi^2$.   Nonperturbatively - in the dual theory -   there are productions of massive gauge 
bosons  of  the broken $SU(N)/U(1)^{N-1}$ generators,    which mix  different $  U(1)^{N-1}$ vortices
and  eventually  lead   to the unique stable vortex  with  a given ${\cal N}$-ality.   There  seem to be no general 
reasons to believe that the tension ratios  found in the small $m$  limit  not be renormalized in such processes.

   Below   we report the result on  the first type of effects: perturbative corrections  to the tension
ratios Eq.(\ref{sineform}), due to the next-to-lowest contributions in $m $.  We shall find a small non-universal 
correction to the sine formula Eq.(\ref{sineform}).     Our point is of course   not that such a 
result is of interest in itself as a physical prediction but that it gives a strong indication for  the non-universality  
of this formula, even though  it could be an approximately a good one.

 The problem of the next-to-lowest contributions in $m $ has been already  studied in
$SU(2)$ theory, by Vainshtein and Yung \cite{YUNG} and by Hou \cite{HOU}, although in that case  there is only one $U(1)$ factor so that 
the author's  interest was different.      When
only up to the order
$A_D$  term in the expansion
\beq m \,    \bra \Tr\,  \Phi^2 \ket  =  m \,  U (A_D)  =  m \, \Lambda^2 ( 1 -  { 2 i A_D \o \Lambda}  -  { 1\o 4}  { A_D^2 \o \Lambda^2} + \ldots  ) 
 \eeq   
is kept,   the effective low energy theory turns out to be  an   ${\cal N}=2$   SQED,  $ A_D$  being an ${\cal N}=2$ analogue  of 
the Fayet-Iliopoulos term.  As a result,   the vortex remains  BPS-saturated, and its  tension is proportional to the 
 monopole charge.   
When  the $ A_D^2$  term is 
taken into account,  the vortex   ceases to be BPS-saturated:  the correction to the vortex tension can be calculated
perturbatively, giving rise to the results  that the vacuum behaves as a type I superconductor.   

\noindent {\bf 2. }   Our aim here is  to generalize the analysis of Vainshtein, Yung and Hou \cite{YUNG,HOU}  to $SU(N)$ theory.   In fact,  
Douglas-Shenker result Eq.(\ref{sineform})  in $SU(N)$ theory  was obtained   in the BPS approximation, by keeping only  the  linear
terms  in $a_{Di}$    in  the expansion 
\beq   \,  U (a_{Di})   =   U_0 +  U_{0k} \, a_{Dk} +  { U_{0 mn}\o 2}   \, a_{Dm}\,  a_{Dn} + \ldots,  
\qquad     U_{0k} = - 4 \,i  \,\Lambda \sin{ \pi k \o N}.  
\eeq
The coefficients $U_{0k}$ were computed by Douglas-Shenker \cite{DS}.
Our first task is then to compute the coefficients of the second term  $U_{0 mn}$.  In principle it is a straightforward matter, 
as one must  simply  invert  the Seiberg-Witten formula:\footnote {We follow  the notation of \cite{DS},  
with   $y^2=  P(x)^2  - \Lambda^2$;   $  P(x)=  { 1\o 2}   \prod_{i=1}^N (x - \phi_i)$   } 
\beq 
a_{Dm} = \oint_{\alpha_m} \lambda, \qquad a_{m} = \oint_{\beta_m} \lambda,\qquad 
\lambda =  { 1 \o 2 \pi i} { x \o y}  { \de P (x) \o \de x} dx, 
\eeq
 which is explicitly known, to second order.  The only trouble is that  $ a_{Dm} $ and $a_{m} $ ($m=1,2,\ldots, N-1$)    are given simply  in
terms of $N$  dependent vacuum parameters
$\phi_i, $   $\sum_{i=1}^N \phi_i =0$.    By denoting the formal derivatives with respect to $\phi_i$ as  ${ \delta \o \delta \phi_i}$,  one finds 
\beq   \sum_{i=1}^N  {\delta   a_{Dm} \o \delta \phi_i } { \de \phi_i \o \de \,a_{Dn}   } =   \delta_{mn}, \qquad 
\sum_{m=1}^{N-1}   { \de \phi_i \o \de \,a_{Dm}   }  {\delta   a_{Dm} \o \delta \phi_j } =   \delta_{ij} - { 1\o N}, 
\label{rela} \eeq
which follow easily by using the constraint,  $\sum_{i=1}^N \phi_i =0$.  In terms of 
$ B_{mi}\equiv  -i  { \delta  a_{D m} \o \delta \phi_i}$,   $  A_{mi}\equiv  -i  { \delta  a_{m} \o \delta \phi_i}$   which are 
explicitly given at the $N$ confining vacua in \cite{DS},    one then finds   
\beq   {\de \phi_i  \o \de \, a_{D m} } = - i B_{mi};  \qquad   \sum_{i=1}^N B_{mi} B_{ni} = \delta_{mn}; \qquad 
\sum_{m=1}^{N-1}  B_{mi} B_{mj}  = \delta_{ij} - { 1\o N}.
\eeq
The explicit values of $B_{mi}$ are (see \cite{DS}):
\begin{equation} 
 B_{mi}= \frac{1}{N} \frac{ \sin[\widehat{\theta}_m ]} {\cos [\theta_i] - 
\cos[\widehat{\theta}_m]  }; \qquad \hat{\theta}_n=\frac{\pi n}{N};  \qquad
\theta_n = \frac{\pi (n-1/2)}{N}.  \end{equation} 
The definition of $u(a_{Di})$ is the following:
\beq u(a_{Di})=\sum_i \phi_i^2. \eeq
Then the desired coefficients can be found by the following expression,
computed at  $a_{Di}=0$: 
\begin{equation} \label{grossa} U_{0mn}=\frac{\partial
^2 u}{\partial a_{Dm}
\partial a_{Dn}} = 2 \sum_k    
 \frac{\partial \phi_k}{\partial a_{Dm}} \frac{\partial \phi_k}{\partial
a_{Dn}} + 2 \phi_k \frac{\partial ^2 \phi_k}{\partial a_{Dm} \partial
a_{Dn}}.    
\end{equation} 
The first part of Eq.(\ref{grossa}) becomes:
\begin{equation} 2 \sum_k
 \frac{\partial \phi_k}{\partial a_{Dm}} \frac{\partial \phi_k}{\partial
a_{Dn}}=- 2 \sum_k B_{km} B_{kn}= - 2 \sum_{k,s} \frac{2}{N}\sin \left[
 \frac{\pi m s}{N} \right]
  \sin \left[
 \frac{\pi n k}{N} \right] \delta_{ks} = - 2 \delta_{mn}.
 \end{equation}
The evaluation of the second term in Eq.(\ref{grossa}) is reported in appendix A. The result is the
following:
\beq 2 \sum_k \phi_k \frac{\partial ^2 \phi_k}{\partial a_{Dm}
 \partial a_{Dn}} = \left( 2-\frac{1}{N} \right) \delta_{mn},
\label{calcolone} \eeq 
thus
 \beq       U_{0 mn}  =  (-   { 1\o N} )  \, 
\delta_{mn}. \label{result} \eeq


\noindent {\bf 3. }  We now use this  result to calculate the corrections to the tension ratios  (\ref{sineform})   found in the lowest order. 
The effective  Lagrangean  near one of the $N$   confining ${\cal N}=1$ vacua    is 
\begin{eqnarray} \label{lagry}
\lefteqn{\mathcal{L} = \sum_{i=1} ^{N-1} Im \left[
\frac{i}{e_{Di}^2} \left( \int d^4 \theta A_{Di} A_{Di}^+ + \int
d^2 \theta (W_{Di})^2 \right)      
\right]+ {} } \nonumber\\
 & & {} +Re \left[\int d^4 \theta (M_i^+ e^{V_{Di}} M_i
+\tilde{M}_i^+ e^{-V_{Di}} \tilde{M}_i)\right] + {} \nonumber\\
 & & {} + 2 Re \left[ \sqrt{2} \int d^2 \theta A_{Di} M_i \tilde{M}_i
+ m   \,  U[A_{Di}]
 \right].  
\label{lagr}  \end{eqnarray}
The coupling constant ${e_{Di}^2} $   is formally vanishing, as 
\[  \frac{4 \pi}{e^2_{Dk}}\simeq \frac{1}{2 \pi} \ln \frac{\Lambda
\sin(\widehat{\theta}_k)}{a_{Dk} N} \]    where  $\widehat{\theta}_m\equiv { \pi n \o N}  $  and  
$a_{Dk}=0$ at the minimum. Physically, the monopole loop integrals
are in fact cut off by masses caused  by the $\mathcal{N}=1$ perturbation.
The monopole  becomes massive  when  $ m \ne 0$, and   $\sqrt2 a_{Dk}$ should be replaced
by the physical monopole mass   $ (m\Lambda \sin( \widehat{
\theta}_k))^{1/2}$  which acts as the infrared cutoff for the coupling constant evolution. 
This is equivalent to the prescription of taking 
$a_{Dn}=<M \tilde{M}>_n^{1/2}$, which is used in \cite{DS}.
 One finds thus 
\begin{equation} \label{accoppiamenti}
e^2_{Dm}\simeq \frac{16 \pi^2}{\ln(\frac{\Lambda \sin(
\widehat{\theta}_m)}{ m \,  N^2})}.
\label{effectivee}  \end{equation}

As $ U_{0 mn}  $ is found to be diagonal,  the description of the ANO vortices \cite{ABR,NO}  in terms of effective magnetic Abelian theory
description continues to be valid for each $U(1)$ factor.    In the linear
approximation $U(A_D) =  m \Lambda^2  +  \mu   A_D, $  where  $\mu \equiv | 4
\,  m  \,\Lambda \sin{ \pi k \o N}|$  for the $k$-th   $U(1)$ theory,  the
theory can be (for the static configurations)    effectively reduced  to an
${\cal N}=4$   theory in $2+1$ dimensions. In this way, Bogomolny's equations
for the BPS vortex can be easily found from the condition that the vacuum to
be supersymmetric: \begin{equation} F_{12}=\sqrt{2}\, (\sqrt{2}M^+ \tilde{M}^+
- \mu ) \qquad  (D_1 + i D_2) M = 0 \end{equation} \begin{equation}
M=\tilde{M}^+,  \qquad    A_D=0.  \end{equation} The solutions of these
equations are similar to the one considered by Nielsen and Olesen: 
\begin{equation} M=\left( \frac{\mu}{\sqrt{2}} \right) ^{1/2} e^{i n \phi} f[r
e \sqrt{\mu}], \qquad  A_{\phi}=-2 n \frac{g(r e \sqrt{\mu} )}{r}
\label{unperturb} \end{equation}
where 
\beq  f'=\frac{f}{r}(1-2g)) n \qquad 
g'=\frac{1}{2 n} r (1-f^2)  \eeq
with boundary consitions  $ f(0)=g(0)=0 $, $f(r\rightarrow
\infty)=1$, $g(r \rightarrow \infty)=+1/2 $). The tension  turns out to be independent of the coupling constant: for the minimum 
vortex \footnote{The fact that the absolute value of   $m$ appears in Eq.(\ref{tension}), as it should, may not be obvious. 
This can be shown by an  appropriate redefinition of the field variables, used in \cite{DPK},  which renders  all equations real.  
The correction term in (\ref{Result})  is  thus negative independently of the phase of $m$.  }
 \beq T=\sqrt{2}
\pi \mu =  4 \sqrt{2} \pi \,  |m  \,\Lambda|  \sin{ \pi k \o N}. \label{tension} \eeq

When the second order term in $U(A_D)= \mu A_D+\frac{1}{2} \eta A_D^2$,  $\eta \equiv U_{kk},$  is taken into account,  
the vortex ceases to
be BPS saturated.   The corrections to the vortex  tension  
due to $\eta$  can be taking into account by perturbation theory,  following  \cite{HOU}.  To first order, the equation for $A_{Dk}=A_D$ is 
\begin{equation}
\nabla^2 A_{D}= - 2 e^4 \eta \, (\mu -\sqrt{2} M \tilde{M}) + 2
e^2 A_{D}  (M M^+ + \tilde{M} \tilde{M}^+)
\end{equation}
where unperturbed expressions  from Eq.(\ref{unperturb}) can be used for $M$, $\tilde{M}. $  The vortex tension becomes simply
\begin{equation}
T=\int d^2 x \, [ \, (-\sqrt{2} \mu F_{12})-2 e^2 \eta A_D (\mu - \sqrt{2}
M^+ \tilde{M}^+) \, ]
\end{equation}
where the second term represents the correction.   By restoring the $k$ dependence, we finally get for the 
tension of the $k$-th   vortex, 
\begin{equation} \label{mipiace}
T_k =4 \sqrt{2}\, \pi \,  |m| \, \Lambda \sin \left(  \frac{\pi k}{N} \right)
-C \frac{   16 \pi^2   |m|^2 }{N^2 \ln \frac{\Lambda \sin(k \pi /N)} {|m|  \, 
N^2 }},  \label{Result} 
\end{equation}
where  $C=2 \sqrt{2} \pi (0.68)=6.04. $   The correction term  has a negative sign, independently of the phase of the adjoint mass. 
 Note that the relation $T_k = T_{N-k}$  continues to  hold.  
        Eq.(\ref{Result})  is valid for $m \ll \Lambda$.

We end with a few remarks.   In the above  consideration,   we have taken into account exactly the $m^2 $ corrections in the F-term of the
effective low-energy action.  On the other hand,  the corrections to the D-terms are   subtler.   Indeed,
based on the physical consideration,   $a_D$  in the argument of the logarithm in the effective low energy coupling constant was replaced by the
monopole mass, of  $O(\sqrt {m \Lambda}).$  This amounts to the $m$ insertion to all orders in the loops.  Such a resummation is necessitated by
the infrared divergences, just as in the case of chiral perturbation theory.    This explains the non-analytic dependence on $m$ as
well as on 
${ 1\o N}$  \cite{Fer}.    

  Also, there are  corrections due to nondiagonal elements  in the coupling constant matrix 
$\tau_{ij}$, which mix the different $U(1)$
factors
\cite{Edel},  neglected in Eq.(\ref{lagr}).  These nondiagonal elements  are suppressed by  $O({1 \o \log{\Lambda/m}})$ relatively to the diagonal 
ones,  apparently of the same order of suppression as the correction calculated  above. However,   these nondiagonal elements
gives rise to terms of the form to the effective potential \cite{Edel}
\bea   \Delta V  &=&   (\Im \tau )^{-1}_{ij}  \,\left(\sqrt 2 M_i \tilde{M}_i -m   {\de   U[A_{D}] \o \de A_{Di} } \, 
\right)^{*}  
\left(\sqrt 2 M_j \tilde{M}_j - m   {\de   U[A_{D}] \o \de A_{Dj} } \right)   +   \non   \\
&+&   {  (\Im \tau )^{-1}_{ij}  \o 2}   \left(     |M_i|^2  -   |\tilde{M}_i|^2  \right )   \left(     |M_j|^2  -   |\tilde{M}_j|^2 \right). 
 \eea
When this is used in  the equations of motion,  one finds that 
 the corrections  to  the tension  due to the nondiagonal  $(\Im \tau )^{-1}_{ij} $  is  actually of one order higher,  $O({1 \o
\log^2{\Lambda/m}}),$ hence is negligible to the order considered. 

We thus find a non-universal  correction to the Douglas-Shenker formula,  Eq.(\ref{sineform}).   In the process of 
transition towards fully non-Abelian superconductivity  at large $m$   nonperturbative effects such as the $W$ boson productions are 
probably  essential. Nonetheless,    the presence of  a calculable deviation  from the sine formula is qualitatively  significant  and shows
 that such a ratio is not a universal  quantity.

   \section* { Acknowledgments  }  

We thank  M. Campostrini, L. Del Debbio, P. Di Vecchia, H. Panagopoulos and  E. Vicari    for useful discussions.  All numerical and algebraic
analysis have been done by using   Mathematica 4.0.1  (Wolfram Research).

\appendix
\section{Computation of (2.\ref{calcolone})}
We use the following identity  (found  by partial derivation of the
identity  $ \frac{\partial a_{Dk} }{\partial a_{Dm} }=\delta_{km}$ - the
first of Eq.(\ref{rela}) - with respect to $a_{Dl}$): \[
  \sum_{i,j} \frac{\partial \phi_{j}}{\partial a_{Dl}}
   \frac{\partial \phi_i}{\partial a_{Dm}}
    \frac{\delta ^2 a_{Dk} }{\delta \phi_{j} \delta \phi_{i}} = -
    \sum_{t} \frac{\delta a_{Dk}}{\delta \phi_{t}}
    \frac{\partial ^2 \phi_{t} }{\partial a_{Dm} \partial a_{Dl}}.
\]
The expression (\ref{calcolone}) now becomes:
\begin{equation} \label{intuito}
  \sum_t \phi_t \frac{\partial ^2 \phi_t}{\partial a_{Dm} \partial a_{Dl}} =
-\sum_{i,j,k,t}
 \phi_{t} \frac{\partial \phi_t}{\partial a_{Dk}}
\frac{\partial \phi_j}{\partial a_{Dl}} \frac{\partial
\phi_i}{\partial a_{Dm}}
   \frac{\delta ^2 a_{Dk} }{\delta \phi_j \delta \phi_i}.
\end{equation}
$\frac{\delta ^2 a_{Dk} }{\delta \phi_j \delta
\phi_i}$ can be found   by first considering  $ \frac{d^2
\lambda}{d \phi_i d \phi_{j}} $ and integrating  this along  the cicles $\alpha_m$
(see \cite{DS} for the conventions; the variable $\theta$ is defined by $x=2
\cos[\theta]$): \begin{equation}
\frac{d^2 \lambda}{d \phi_i d \phi_{j}}=\frac{1}{2 \pi i} \left[
-\frac{P (1-\delta_{ij}) }{y (x-\phi_i)(x-\phi_j)}dx+
\frac{P^3}{y^3(x-\phi_i)(x-\phi_j)}dx \right]=
\end{equation}
\[ =-\frac{(1-\delta_{ij} )\cot(N \theta)
\sin(\theta) d \theta}{\pi (x-\phi_i)(x-\phi_j)} -\frac{\cot^3(N
\theta) \sin(\theta) d \theta}{\pi (x-\phi_i)(x-\phi_j)}. 
\]
We perform the integrations taking the residues at the poles: $\widehat{\theta}_m=\frac{\pi m}{N}$:
\begin{equation}
 \frac{\delta ^2 a_{Dm} }{\delta \phi_i \delta \phi_j}
_{a_{D}=0} =
 +\frac{i}{2} ( Res [A] + Res [B] ),
\end{equation}
\begin{equation}
A=\frac{ \cot[N \theta] \sin[\theta] (1-\delta_{ij} )} {(\cos
[\theta] -
 \cos[\theta_i] )( \cos [\theta] - \cos[\theta_j]) },
\end{equation}
\begin{equation}
B=\frac{ \cot^3[N \theta] \sin[\theta] } {(\cos [\theta] -
 \cos[\theta_i] )( \cos [\theta] - \cos[\theta_j])}.
\end{equation}
At the end of the integration, one has:
\[ \frac{2}{i} \frac{\delta ^2 a_{Dt} }{\delta \phi_i \delta \phi_j}=
\frac{-\sin(\widehat{\theta}_t)}{( \cos [\widehat{\theta}_t] -
\cos[\theta_i])( \cos [\widehat{\theta}_t] - \cos[\theta_j])}
\left( \frac{1}{2 N^3} +\frac{\delta_{ij}}{N} \right) \]
\[ -\frac{1}{4 N^3} \left[ \frac{\sin [\widehat{\theta}_t] (\cos[2\widehat{\theta}_t]
 +2 \cos[\theta_i] \cos[\widehat{\theta}_t]-3) } {( \cos
  [\widehat{\theta}_t] - \cos[\theta_i])^3 ( \cos
[\widehat{\theta}_t] - \cos[\theta_j])} \right] \]
\[ -\frac{1}{4 N^3} \left[ \frac{\sin [\widehat{\theta}_t] (\cos[2\widehat{\theta}_t]
 +2 \cos[\theta_j] \cos[\widehat{\theta}_t]-3) } {( \cos
  [\widehat{\theta}_t] - \cos[\theta_i]) ( \cos
[\widehat{\theta}_t] - \cos[\theta_j])^3} \right] \].
\begin{equation} \label{casino}
 +\frac{1}{2N^3} \left[ \frac{2 \sin^3 [\widehat{\theta}_t] +\sin [2\widehat{\theta}_t]
 ( 2 \cos[\widehat{\theta}_t] -(\cos[\theta_i]+\cos[\theta_j]))  } {( \cos
  [\widehat{\theta}_t] - \cos[\theta_i])^2 ( \cos
[\widehat{\theta}_t] - \cos[\theta_j])^2}  \right]
\end{equation}
Substituting  this in (\ref{intuito})  one finds  
\bea    && 2 \sum_k \phi_k \frac{\partial ^2 \phi_k}{\partial a_{Dm}
 \partial a_{Dn}} = 4 i  \sum_{t,r,s} \sin \left[ \frac{\pi t}{N} \right]
 \frac{\partial \phi_s}{\partial a_{Dn}} \frac{\partial \phi_r}{\partial
a_{Dm}}    \frac{\delta ^2 a_{Dt} }{\delta \phi_r \delta \phi_s} = \non \\
&=&    -4 i  \sum_{t,r,s} \sin \left[ \frac{\pi t}{N} \right] B_{ns}
B_{mr} \frac{\delta ^2 a_{Dt} }{\delta \phi_r \delta \phi_s}
=\left( 2-\frac{1}{N} \right) \delta_{mn}.  \label{sorpresa}
\eea  
The last equality involves rather cumbersome  trigonometric expressions:  we found Eq.(\ref{sorpresa}) by using
Mathematica up to  $N=50$.


\begin{thebibliography}{100}

\bibitem{STR}     A. Hanany, M. Strassler and A. Zaffaroni,  {\bf Nucl.Phys. B513}   (1998) 87,   hep-th/9707244.

\bibitem{Kleb} C. P. Herzog and I. R. Klebanov,  {\bf Phys.Lett. B526} (2002) 388, 
hep-th/0111078. 



\bibitem{LT}  B.  Lucini and  M. Teper,
{\bf Phys.Lett. B501} (2001) 128,   hep-lat/0012025;  
{\bf Phys.Rev.D64} (2001) 105019,
 hep-lat/0107007. 

   

\bibitem{Pisa} L. Del Debbio, H. Panagopoulos, P. Rossi and  E. Vicari,  {\bf Phys.Rev. D65} (2002) 021501, hep-th/0106185;
{\bf JHEP 0201} (2002) 009,  hep-th/0111090  

\bibitem{DS}  M.R. Douglas and  S.H. Shenker,   {\bf Nucl. Phys.  B447}
(1995) 271,   hep-th/9503163.

\bibitem{STRASS}  M. Strassler,   {\bf Progr.  Theor. Phys. Suppl. 131} (1998) 439,  hep-lat/9803009.


\bibitem{YUNG}   A. Yung,   hep-th/0005088, 
3rd Moscow School of Physics and 28th ITEP Winter School of Physics, Moscow, 2000; 
 A. Vainshtein and A. Yung,  {\bf Nucl.Phys.B614},  3,2001,   hep-th/0012250.

\bibitem{TM}  G. 't Hooft, {\bf  Nucl. Phys.   B190}   (1981) 455.
   S. Mandelstam,  {\bf Phys. Lett.  53B }  (1975) 476;
                                {\bf  Phys. Rep.   23C} (1976) 245.
 

\bibitem{ABR}  A.A. Abrikosov, {\bf  JETP  5}  (1957) 1174.

\bibitem{NO}   H. Nielsen and P. Olesen,   {\bf Nucl. Phys. B61}  (1973) 45.


\bibitem{DPK}  M. Di Pierro and K. Konishi, {\bf Phys. Lett. B388} (1996) 90, 
 hep-th/9605178. 


\bibitem{HOU}
X.~r.~Hou,   {\bf Phys. Rev.  D  63} (2001) 045015, hep-th/0005119.

\bibitem{Fer}  F. Ferrari, {\bf Nucl.Phys. B612} (2001) 151,  hep-th/0106192 

\bibitem{Edel}  J.D. Edelstein, W.G. Fuertes,  J. Mas and  J. M. Guilarte, 
{\bf Phys.Rev. D62} (2000)  065008, hep-th/0001184. 

\end{thebibliography}
\end {document}